# Global Attitude Synchronization of Networked Rigid Bodies Under Directed Topologies

Fan Zhang, Deyuan Meng, *Senior Member, IEEE*, and Jingyao Zhang

**Abstract**

The global attitude synchronization problem is studied for networked rigid bodies under directed topologies. To avoid the asynchronous pitfall where only vector parts converge to some identical value but scalar parts do not, multiplicative quaternion errors are leveraged to develop attitude synchronization protocols for rigid bodies with the absolute measurements. It is shown that global synchronization of networked rigid bodies can be achieved if and only if the directed topology is quasi-strongly connected. Simultaneously, a novel double-energy-function analysis method, equipped with an ordering permutation technique about scalar parts and a coordinate transformation mechanism, is constructed for the quaternion behavior analysis of networked rigid bodies. In particular, global synchronization is achieved with our analysis method regardless of the highly nonlinear and strongly coupling problems resulting from multiplicative quaternion errors, which seriously hinder the traditional analysis of global synchronization for networked rigid bodies. Simulations for networked spacecraft are presented to show the global synchronization performances under different directed topologies.

## I. Introduction

Attitude synchronization of networked rigid bodies plays a critical role in many applications, such as multiple quadrotors [1], [2], spacecraft formation flying [3]–[5], and multiple-robot systems [6]. This problem is generally solved with the design of synchronization protocols for

The authors are with the Seventh Research Division, Beihang University (BUAA), Beijing 100191, China, and also with the School of Automation Science and Electrical Engineering, Beihang University (BUAA), Beijing 100191, China (email: zhangfan_buaa@buaa.edu.cn, dymeng@buaa.edu.cn).

networked rigid bodies. As reported in [7], networked rigid bodies can be driven to an identical orientation, priori unknown or prescribed, with the synchronization protocols based on attitude information of its nearest neighbors and itself. One of the most fundamental problems for protocol design and analysis is that the attitude configuration space is not diffeomorphic to a Euclidean space [8]. In consideration of this problem, most existing protocols are proposed by attitude representations such as Euler angles [9] and Rodrigues Parameters [10]. It is worth mentioning that both representations are either kinematically or geometrically singular, which are limited to the local attitude maneuvers [11]. Modified Rodrigues Parameters can be employed to represent all physical orientations, but the singularity is inevitable when the rotation angle is $\pm 2\pi$ [12]. The axis with angle zero should be carefully defined for the axis-angle representation [13]. In addition, while the rotation matrices can be considered as one approach to cover all physical attitudes, not all the information available in one rotation matrix are necessary for the attitude representation [11]. To avoid the inherent singularities and use as few parameters as possible, we consider quaternions as a good alternative to map all physical attitudes.

In the case of a quaternion, four parameters satisfying the unity constraint are generally applied to represent the physical orientations of rigid bodies. As introduced in [1], quaternions do not give rise to the aforementioned singularities. However, there are two basic issues when quaternions are used. The first issue is that quaternions double cover SO(3), which is called the special orthogonal group [5], that is, one physical attitude always corresponds to a pair of antipodal quaternions [11]. An implication of this nonunique projection is that properties for kinematics of rigid bodies on SO(3) can not be derived using quaternions directly. Therefore, the quaternion representation should be treated carefully to escape from the abovementioned two-fold covering phenomenon. The second issue is how to construct attitude errors between two rigid bodies after obtaining *absolute* orientation information [10]. In the literature, some protocols with additive quaternion errors are developed [4], [14]–[16]. Of note is that these protocols may give rise to the non-unity constraint [4] or an asynchronous pitfall where only vector parts converge to some identical value while the scalar parts do not [14]–[16]. The other approach to overcome this shortfall involves multiplicative quaternion errors [17]–[19]. Therefore, the development of multiplicative quaternion error-based synchronization protocols is well-motivated.

Nevertheless, the analysis of the global objective based on multiplicative quaternion errors is quite challenging, and one of main difficulties lies in that multiplicative quaternion errors involve strong coupling problem arising from the product of the scalar parts and the vector parts of quaternions. To address this problem and thus reach synchronization, some progress has been made in the recent past, while most analysis methods are only applicable to the circumstances under some restrictive initial conditions. In [10], the proposed analysis method is only applied to the condition that the initial attitudes are contained in a positively invariant set. In [18], all the initial orientation matrices should be assumed to be positive definite. From [19], the scalar parts of quaternions should be positive at the initial moment. Note that the initial orientation restriction is helpful to handle the aforementioned coupling problem in traditional analysis frameworks, but it hinders the achievement of global attitude synchronization. Hence, it is of great significance to develop an appropriate analysis method, with which the initial orientation restriction can be relaxed and thus the global objective can be achieved.

Moreover, it is worth mentioning that most of the existing works based on multiplicative attitude errors focus on solving synchronization problems under the undirected topologies or the strongly connected topologies. In [19], the synchronization can be reached only if the communication graph is restricted to be strongly connected, because the positive left eigenvector of the Laplacian matrix associated with the eigenvalue zero does not exist under quasi-strongly connected topologies. For the analysis of [17] and [20], the symmetric property of the weighted adjacency matrix of an undirected graph is utilized to obtain the attitude synchronization. The strict connectivity condition of the topologies, obviously, can not be guaranteed sometimes because of the complexity of the communication environment, and thus it is required that attitude protocols can be applied to quasi-strongly connected topologies. Hence, how to implement the synchronization analysis for quasi-strongly connected topologies is another task of interest.

Motivated by the task of almost global synchronization with the absolute attitude measurement in [10], also investigated in [3], [17]–[19], [21], we are aimed at *global* attitude synchronization of networked rigid bodies using absolute attitude information. Further, a necessary and sufficient condition on topologies is obtained for global synchronization. The main contributions of this paper are summarized as follows.

1) To handle the coupling problem arising from multiplicative quaternion errors and achieve attitude synchronization under directed topologies, a double-energy-function analysis method, relying on no eigenvectors of Laplacian matrix, is firstly developed. It is verified that attitudes of all rigid bodies can eventually converge to some identical attitude if not all scalar parts of initial quaternions are zero, which replaces the initial condition with all positive scalar parts of initial quaternions should be positive in [10], [19], [21].

2) To further relax the initial orientation constraints, a novel coordinate transformation-based analysis method is proposed. Especially for quasi-strongly connected topologies, it is difficult to prove the existence of appropriate transformations. Thanks to this method, attitude synchronization can be achieved regardless of the initial orientations, which implies that the initial condition in [17]–[19] can be relaxed to the global scope;
3) It is concluded that global attitude synchronization can be accomplished *if and only if* topologies are quasi-strongly connected, which contains the case of strong connect. As a result, the application of networked rigid bodies with the absolute attitude measurement is further expanded. This can not be achieved in the existing works, e.g., [17], [19], [20], where the proposed methods heavily rely on the strong connectivity of topologies or symmetric property of the weighted adjacency matrix.

The remaining parts of this paper are organized as follows. In Section II, some useful preliminaries for graph theory and quaternion-based attitude kinematics are firstly introduced, followed by the problem formulation and the protocol design in Section III. The synchronization over strongly connected topologies and quasi-strongly connected topologies are presented in Sections IV and V, respectively. In Section VI, simulations under different directed topologies are provided. Conclusions are drawn in Section VII.

*Notations:* For a matrix $\boldsymbol{A} = [a_{ij}] \in \mathbb{R}^{m\times n}$, $||\boldsymbol{A}||$ denotes the spectral norm of $\boldsymbol{A}$. Let $n = 1$, and then $||\boldsymbol{A}||$ denotes the $l_2$ norm of $\boldsymbol{A} \in \mathbb{R}^m$. $\boldsymbol{0}$ and $\boldsymbol{I}$ denote the null matrix and the identity matrix with required dimensions, respectively. $\mathbb{S}^2 = \{\boldsymbol{x} \in \mathbb{R}^3 : ||\boldsymbol{x}|| = 1\}$ is the two-sphere and $\mathbb{S}^3 = \{\boldsymbol{x} \in \mathbb{R}^4 : ||\boldsymbol{x}|| = 1\}$ is the three-sphere. For any $\boldsymbol{x} = [x_1, x_2, x_3]^{\mathrm{T}} \in \mathbb{R}^3$, the vector cross-product operator is defined as

$$\boldsymbol{x}^{\times} = \begin{bmatrix} 0 & -x_3 & x_2 \\ x_3 & 0 & -x_1 \\ -x_2 & x_1 & 0 \end{bmatrix},$$

which shows $||\boldsymbol{x}^{\times}|| = ||\boldsymbol{x}|| = \left|\left|\boldsymbol{x}^{\mathrm{T}}\right|\right|$ and $\boldsymbol{x}^{\mathrm{T}}\boldsymbol{x}^{\times} = \boldsymbol{0}$.

## II. Preliminaries

In this section, some concepts about graph theory are firstly reviewed. Then the quaternion-based kinematics is introduced to facilitate the attitude synchronization protocol developed in the subsequent sections.

### A. Graph theory

The communication topology is described as a directed graph denoted by $\mathscr{G} = \{\mathscr{N}, \mathscr{E}\}$, where $\mathscr{N} = \{1, \ldots, N\}$ is the node set and $\mathscr{E} \subseteq \mathscr{N} \times \mathscr{N}$ is the directed edge set. A directed edge $(j, i)$ denotes that $j$ is a neighbor of $i$ such that $i$ can directly receive the information from $j$. The neighboring nodes of $i$ constitute a set $\mathscr{N}_i = \{j : j \in \mathscr{N} \text{ and } (j, i) \in \mathscr{E}\}$. The weighted adjacency matrix is defined as $\mathscr{A} = [a_{ij}] \in \mathbb{R}^{N \times N}, i, j \in \mathscr{N}$ such that $a_{ij} > 0$ if $(j, i) \in \mathscr{E}$ and $a_{ij} = 0$, otherwise. It is worth pointing that $(i, i) \notin \mathscr{E}$ and thus $a_{ii} = 0$, $\forall i \in \mathscr{N}$. In particular, $\mathscr{G}$ collapses into an undirected graph if $\mathscr{A} = \mathscr{A}^{\mathrm{T}}$. The degree matrix is defined as $\mathscr{D} = \operatorname{diag}\{d_i\} \in \mathbb{R}^{N \times N}$ where $d_i = \sum_{j \in \mathscr{N}_i} a_{ij}$. The Laplacian matrix of $\mathscr{G}$, denoted by $\mathscr{L} \in \mathbb{R}^{N \times N}$, satisfies $\mathscr{L} = \mathscr{D} - \mathscr{A}$. A path is formed by a finite sequence of edges in $\mathscr{E}$. Moreover, if, for certain node $i$, there exist paths from $i$ to every other node in $\mathscr{G}$, then we have that $\mathscr{G}$ is quasi-strongly connected and $i$ is a root. Especially, $\mathscr{G}$ is also said to be strongly connected if all nodes in $\mathscr{G}$ are roots [22]. For example, Figs. 1-2 intuitively show the strong connectivity and the quasi-strong connectivity with five rigid spacecraft, which can be used for simulation verification in Section VI.

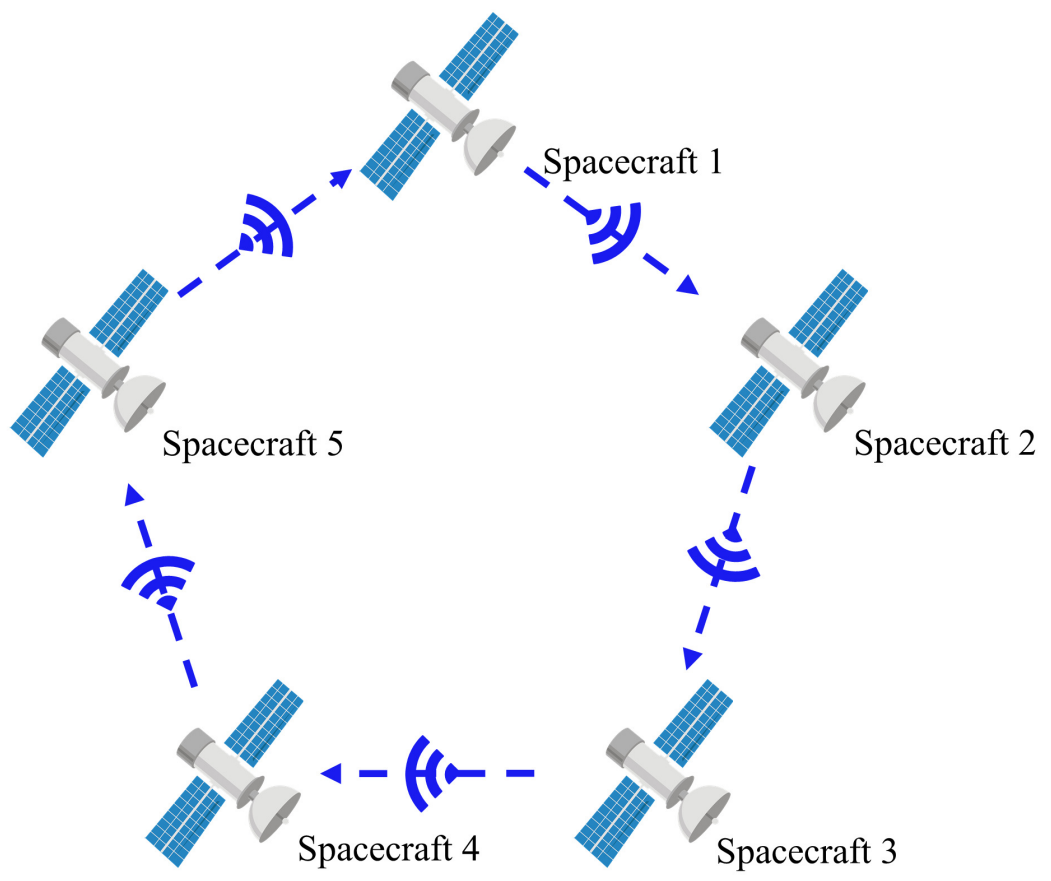

Fig. 1. A strongly connected topology.

### B. Attitude kinematics

In practical applications, the quaternion representation is developed based on Euler's rotation theorem, which states that the relative attitude between any two coordinate frames can be presented as only one rotation about some fixed axis. Generally, the quaternion is treated as $[\varepsilon_i(t), \boldsymbol{q}_i^{\mathrm{T}}(t)]^{\mathrm{T}} \in \mathbb{S}^3$ with $\varepsilon_i(t) \in \mathbb{R}$ and $\boldsymbol{q}_i(t) \triangleq [q_i^{(1)}(t), q_i^{(2)}(t), q_i^{(3)}(t)]^{\mathrm{T}} \in \mathbb{R}^3$ denoting the scalar part and vector part of the quaternion, respectively. According to Euler's rotation theorem, $\varepsilon_i(t)$ and $\boldsymbol{q}_i(t)$ can be expressed as

$$\varepsilon_i(t) = \cos\left(\frac{\phi_i(t)}{2}\right) \text{ and } \boldsymbol{q}_i(t) = \boldsymbol{e}_i(t)\sin\left(\frac{\phi_i(t)}{2}\right), \tag{1}$$

where $\phi_i(t) \in \mathbb{R}$ denotes the Euler angle and $\boldsymbol{e}_i(t) \in \mathbb{R}^3$ denotes the so-called Euler unit axis. From (1), $[\varepsilon_i(t), \boldsymbol{q}_i^{\mathrm{T}}(t)]^{\mathrm{T}}$ and $[-\varepsilon_i(t), -\boldsymbol{q}_i^{\mathrm{T}}(t)]^{\mathrm{T}}$ represent the same physical attitude.

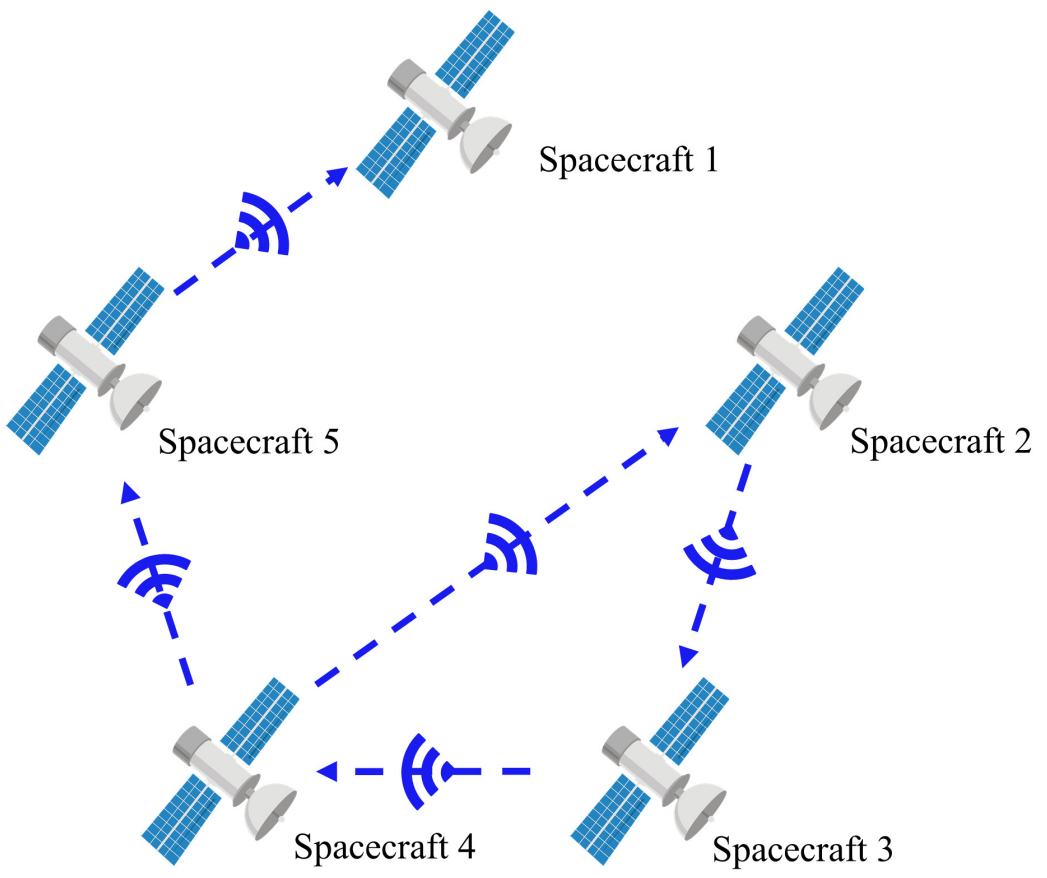

Fig. 2. A quasi-strongly connected topology.

For a group of $N$ rigid bodies, attitude kinematics of rigid body $i \in \mathscr{N}$ governed by Euler's rotational equations of motion is given by

$$\begin{aligned} \dot{\varepsilon}_i(t) &= -\frac{1}{2}\boldsymbol{q}_i^{\mathrm{T}}(t)\boldsymbol{\omega}_i(t) \\ \dot{\boldsymbol{q}}_i(t) &= \frac{1}{2}\left(\boldsymbol{q}_i^{\times}(t) + \varepsilon_i(t)I\right)\boldsymbol{\omega}_i(t), \end{aligned} \tag{2}$$

where $\boldsymbol{\omega}_i(t)$ denotes the angular velocity of rigid body $i$ with respect to the inertial frame $\mathscr{F}_I$ expressed in the body-fixed frame $\mathscr{F}_B^i$. Moreover, from the definition of $\mathbb{S}^3$, we can know that $\varepsilon_i(t)$ and $\boldsymbol{q}_i(t)$ satisfy the unity constraint:

$$\varepsilon_i^2(t) + \boldsymbol{q}_i^{\mathrm{T}}(t)\boldsymbol{q}_i(t) = 1, \tag{3}$$

which implies $|\varepsilon_i(t)| \leqslant 1$ and $\|\boldsymbol{q}_i(t)\| \leqslant 1$. For more introduction and properties of quaternions, the readers are referred to Section 2.2 in [23].

## III. Problem formulation

Considering the preliminaries, we can present the objective as follows. Let the graph $\mathscr{G}$ be quasi-strongly connected and initial physical attitudes of networked rigid bodies be arbitrary. The objective of this work is to develop an analysis method, with which the adopted attitude protocol can be shown to achieve global attitude synchronization, namely,

$$\begin{aligned} \lim_{t\to\infty} |\varepsilon_i(t) - \varepsilon_j(t)| &= 0 \\ \lim_{t\to\infty} \left\|\boldsymbol{q}_i(t) - \boldsymbol{q}_j(t)\right\| &= 0, \end{aligned} \tag{4}$$

for all $i, j \in \mathscr{N}$.

To reach the objective (4), we first introduce the multiplicative quaternion error of rigid body $i$ with respect to rigid body $j$, which can be defined as

$$\begin{aligned} \varepsilon_{i,j}(t) &\triangleq \varepsilon_i(t)\varepsilon_j(t) + \boldsymbol{q}_i^{\mathrm{T}}(t)\boldsymbol{q}_j(t) \\ \boldsymbol{q}_{i,j}(t) &\triangleq \varepsilon_j(t)\boldsymbol{q}_i(t) - \varepsilon_i(t)\boldsymbol{q}_j(t) + \boldsymbol{q}_i^{\times}(t)\boldsymbol{q}_j(t), \end{aligned} \tag{5}$$

where $\varepsilon_{i,j}(t)$ and $\boldsymbol{q}_{i,j}(t)$ denote the scalar part and the vector part of the multiplicative quaternion error, respectively. From [23], $\varepsilon_{i,j}^2(t)+\boldsymbol{q}_{i,j}^{\mathrm{T}}(t)\boldsymbol{q}_{i,j}(t) = 1$ holds. Based on this definition, we adopt the attitude synchronization protocol as

$$\boldsymbol{\omega}_i(t) = -\sum_{j\in\mathscr{N}_i} a_{ij}\boldsymbol{q}_{i,j}(t). \tag{6}$$

**Remark 1.** *The protocol* (6) *is constructed using the absolute attitude measurement introduced in the first case of [10], and is similar with the attitude synchronization protocols in [18], [19]. It is of note that the implement of the protocol in [18] require that* $|\phi_i(0)| < \frac{\pi}{2}$ *for all rigid bodies, which means that* $\varepsilon_i(0) > \frac{\sqrt{2}}{2}$ *and is also used in [24]. In addition, [19] should meet* $\varepsilon_i(0) > 0$ *for all rigid bodies. These results are all non-global synchronization. As one of main contributions, we can show that global attitude synchronization can be achieved under* (6) *with the different analysis method.*

To implement the attitude synchronization protocol (6), we first construct the unique mapping between the physical attitude and the quaternion representation. According to the definition of the initial quaternion representation of rigid body $i$, $\mathbb{S}^3$ can be divided into four different subspaces as

$$
\begin{aligned}
&S_1 \triangleq \left\{ \left[\varepsilon_i(0), \boldsymbol{q}_i^{\mathrm{T}}(0)\right]^{\mathrm{T}} \in \mathbb{S}^3 : \varepsilon_i(0) > 0 \right\} \\
&S_2 \triangleq \left\{ \left[\varepsilon_i(0), \boldsymbol{q}_i^{\mathrm{T}}(0)\right]^{\mathrm{T}} \in \mathbb{S}^3 : \varepsilon_i(0) = 0 \text{ and } \boldsymbol{q}_i(0) \in S_+ \right\} \\
&S_3 \triangleq \left\{ \left[\varepsilon_i(0), \boldsymbol{q}_i^{\mathrm{T}}(0)\right]^{\mathrm{T}} \in \mathbb{S}^3 : \varepsilon_i(0) < 0 \right\} \\
&S_4 \triangleq \left\{ \left[\varepsilon_i(0), \boldsymbol{q}_i^{\mathrm{T}}(0)\right]^{\mathrm{T}} \in \mathbb{S}^3 : \varepsilon_i(0) = 0 \text{ and } \boldsymbol{q}_i(0) \in S_- \right\},
\end{aligned} \tag{7}
$$

where $S_+$ and $S_-$ are defined as

$$
\begin{aligned}
S_+ \triangleq &\left\{ \boldsymbol{x} = [x_i] \in \mathbb{S}^2 : x_3 > 0 \right\} \\
&\cup \left\{ \boldsymbol{x} = [x_i] \in \mathbb{S}^2 : x_3 = 0 \text{ and } x_2 > 0 \right\} \\
&\cup \left\{ \boldsymbol{x} = [x_i] \in \mathbb{S}^2 : x_3 = x_2 = 0 \text{ and } x_1 > 0 \right\} \\
S_- \triangleq &\left\{ \boldsymbol{x} = [x_i] \in \mathbb{S}^2 : \boldsymbol{x} \notin S_+ \right\},
\end{aligned} \tag{8}
$$

respectively and are illustrated in Fig. 3. From the definition of the quaternion in (1), $S_1$ and $S_3$ represent the same physical attitude space, and $S_2$ and $S_4$ represent another same physical attitude space, which means that the entire physical attitude space can be represented by using only $S_1$ and $S_2$. As a result, it is reasonable for us to construct the one-to-one mapping between the physical attitude space and $S_1 \cup S_2$ at the initial time, that is,

$$
[\varepsilon_i(0), \boldsymbol{q}_i^{\mathrm{T}}(0)]^{\mathrm{T}} \in S_1 \cup S_2, \quad \forall i \in \mathscr{N}. \tag{9}
$$

Moreover, to analyze global attitude synchronization under the quasi-strongly connected topologies, we first present following definitions for the discussion. $\mathscr{G} = \{\mathscr{N}, \mathscr{E}\}$ is considered as a quasi-strongly connected topology. Let $\mathscr{N}^{\mathrm{r}} \subseteq \mathscr{N}$ be the root-node set which consists of all root nodes of $\mathscr{G}$. Without loss of generality, we assume that the set $\mathscr{N}^{\mathrm{r}}$ is given by $\mathscr{N}^{\mathrm{r}} = \{1, \ldots, N^{\mathrm{r}}\}$ with $1 \leqslant N^{\mathrm{r}} \leqslant N$. Denote the non-root node set $\mathscr{N}^{\mathrm{nr}} \triangleq \mathscr{N} \backslash \mathscr{N}^{\mathrm{r}} = \{j : j \in \mathscr{N} \text{ and } j \notin \mathscr{N}^{\mathrm{r}}\}$ which can also be expressed as $\mathscr{N}^{\mathrm{nr}} = \{N^{\mathrm{r}}+1, \ldots, N\}$. Denote also the induced directed graph $\mathscr{G}^{\mathrm{r}} = \{\mathscr{N}^{\mathrm{r}}, \mathscr{E}^{\mathrm{r}}\}$, where $\mathscr{E}^{\mathrm{r}} = \{(j,i) \in \mathscr{E} : i,j \in \mathscr{N}^{\mathrm{r}}\}$. It is worth noticing that $\mathscr{N}^{\mathrm{nr}} = \varnothing$ holds and $\mathscr{G}$ collapses into a strongly connected graph if $N^{\mathrm{r}} = N$. With these definitions, we recall a useful lemma about $\mathscr{G}^{\mathrm{r}}$.

**Lemma 1** [25]. *For a quasi-strongly connected directed graph $\mathscr{G}$, $\mathscr{G}^{\mathrm{r}}$ is a subgraph of $\mathscr{G}$, and furthermore, $\mathscr{G}^{\mathrm{r}}$ is always strongly connected.*

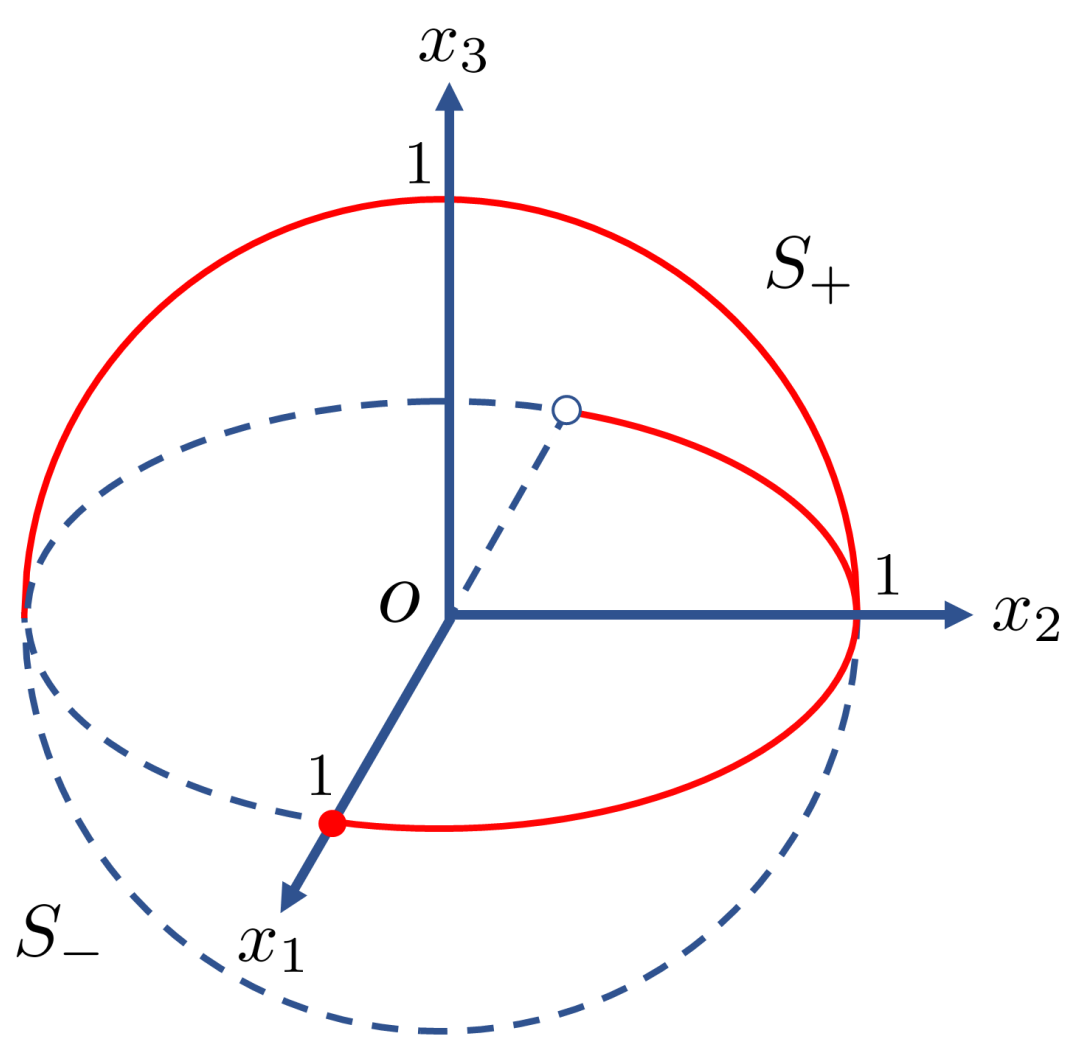

Fig. 3. The illustrations of the sets $S_+$ and $S_-$.

As given in Lemma 1, the strong connectivity of $\mathscr{G}^{\mathrm{r}}$ always holds. In addition, from the definition of the root node, it is obvious that the root nodes would not receive the information from non-root nodes. As a result, it is reasonable for us to start with the analysis of strongly connected topologies in Section IV, which reveals the attitude behaviors of the root nodes, and then move to the analysis of quasi-strongly connected topologies in Section V.

## IV. Synchronization of root-related rigid bodies

In this section, we study the attitude behaviors of the root-related rigid bodies, which can also be considered as a general strongly connected graph described by $\mathscr{G}^{\mathrm{r}} = \{\mathscr{N}^{\mathrm{r}}, \mathscr{E}^{\mathrm{r}}\}$ according to Lemma 1. Based on this fact, we present a double-energy-function analysis method for the evolution of scalar parts and vector parts.

According to (9), we naturally have $[\varepsilon_i(0), \boldsymbol{q}_i^{\mathrm{T}}(0)]^{\mathrm{T}} \in S_1 \cup S_2$, $\forall i \in \mathscr{N}^{\mathrm{r}}$, which can be divided into two conditions as

i1) $\left[\varepsilon_i(0), \boldsymbol{q}_i^{\mathrm{T}}(0)\right]^{\mathrm{T}} \in S_1$, $\exists i \in \mathscr{N}^{\mathrm{r}}$;

i2) $\left[\varepsilon_i(0), \boldsymbol{q}_i^{\mathrm{T}}(0)\right]^{\mathrm{T}} \in S_2$, $\forall i \in \mathscr{N}^{\mathrm{r}}$.

In the following, we focus on the attitude-behavior analysis of the scalar parts of quaternions under i1) and i2), which plays a key role in addressing the coupling problem arising from the product of the scalar parts and the vector parts of quaternions.

## A. Evolution of the scalar parts of quaternions

Before the detailed analysis of the evolution of $\varepsilon_i(t)$, we first recall a useful lemma below about $\min_{i\in\mathscr{N}^{\mathrm{r}}}\varepsilon_i(t)$.

**Lemma 2** [26]. *Let $\varepsilon_*(t)$ be equal to $\min_{i\in\mathscr{N}^{\mathrm{r}}}\varepsilon_i(t)$ and $k(t)$ be the index function such that $\varepsilon_{k(t)}(t)=\varepsilon_*(t)$ for all $t$. In the case of the nonuniqueness of $k(t)$ (i.e., $\varepsilon_i(t)=\varepsilon_j(t)=\varepsilon_*(t)$ for some $i\neq j$ and $t$), there exist some measurable $k(\cdot)$ in a way such that $\dot{\varepsilon}_{k(t)}(t)=\dot{\varepsilon}_*(t)$ for almost all $t$.*

With the existence of the measurable $k(t)$ from Lemma 2, we can propose the following result under the initial condition i1) with the attitude-behavior analysis of the single rigid body $k(t)$ that is generally time-varying.

**Lemma 3.** *Let $\mathscr{G}^{\mathrm{r}}$ be strongly connected. If the initial condition i1) is met and the protocol* (6) *is applied, then, 1) $\dot{\varepsilon}_*(t)\geqslant 0$ for almost all $t$; 2) $\lim_{t\to\infty}\varepsilon_i(t)=C_1,\forall i\in\mathscr{N}^{\mathrm{r}}$ where $C_1$ is a positive constant.*

*Proof. 1)* From Lemma 2, there exists a measurable function $k(t)$ such that $\varepsilon_*(t)=\varepsilon_{k(t)}(t)$ and $\dot{\varepsilon}_*(t)=\dot{\varepsilon}_{k(t)}(t)$ for almost all $t$, regardless of the nonuniqueness of $k(t)$. Thus,

$$\varepsilon_j(t)-\varepsilon_{k(t)}(t)\geqslant 0,\quad \forall j\in\mathscr{N}^{\mathrm{r}}_{k(t)}. \tag{10}$$

To proceed, we can design an energy function as

$$W_1(\varepsilon_{k(t)}(t),\boldsymbol{q}_{k(t)}(t))\triangleq\left\|\boldsymbol{q}_{k(t)}(t)\right\|^2+\left(\varepsilon_{k(t)}(t)-1\right)^2. \tag{11}$$

For simplicity, we denote $W_1(t)=W_1(\varepsilon_{k(t)}(t),\boldsymbol{q}_{k(t)}(t))$ in the subsequent text. Clearly, through $\boldsymbol{q}^{\mathrm{T}}_{k(t)}(t)\boldsymbol{q}^{\times}_{k(t)}(t)=0$ and (2), (11) may directly lead to

$$\begin{aligned}\dot{W}_1(t)&=2\boldsymbol{q}^{\mathrm{T}}_{k(t)}(t)\dot{\boldsymbol{q}}_{k(t)}(t)+2\left(\varepsilon_{k(t)}(t)-1\right)\dot{\varepsilon}_{k(t)}(t)\\&=\boldsymbol{q}^{\mathrm{T}}_{k(t)}(t)\boldsymbol{\omega}_{k(t)}(t)\end{aligned} \tag{12}$$

for almost all $t$. With inserting the attitude synchronization protocol (6) into (12), we have

$$\dot{W}_1(t)=-\sum_{j\in\mathscr{N}^{\mathrm{r}}_{k(t)}}a_{k(t)j}\boldsymbol{q}^{\mathrm{T}}_{k(t)}(t)\boldsymbol{q}_{k(t),j}(t). \tag{13}$$

Considering the expression of $\boldsymbol{q}_{i,j}(t)$ in (5) and invoking $\boldsymbol{q}_{k(t)}^{\mathrm{T}}(t)\boldsymbol{q}_{k(t)}^{\times}(t) = 0$ again, we can transform (13) into

$$\dot{W}_1(t) = -\sum_{j\in\mathscr{N}_{k(t)}^{\mathrm{r}}} a_{k(t)j}\varepsilon_j(t)\boldsymbol{q}_{k(t)}^{\mathrm{T}}(t)\boldsymbol{q}_{k(t)}(t) + \sum_{j\in\mathscr{N}_{k(t)}^{\mathrm{r}}} a_{k(t)j}\varepsilon_{k(t)}(t)\boldsymbol{q}_{k(t)}^{\mathrm{T}}(t)\boldsymbol{q}_j(t). \tag{14}$$

Let us now consider the case of $\varepsilon_{k(t)}(t) \geqslant 0$, it easily follows from (14) that

$$\begin{aligned}\dot{W}_1(t) \leqslant{}& \frac{1}{2}\sum_{j\in\mathscr{N}_{k(t)}^{\mathrm{r}}} a_{k(t)j}\varepsilon_{k(t)}(t)\boldsymbol{q}_{k(t)}^{\mathrm{T}}(t)\boldsymbol{q}_{k(t)}(t) + \frac{1}{2}\sum_{j\in\mathscr{N}_{k(t)}^{\mathrm{r}}} a_{k(t)j}\varepsilon_{k(t)}(t)\boldsymbol{q}_j^{\mathrm{T}}(t)\boldsymbol{q}_j(t)\\ &- \sum_{j\in\mathscr{N}_{k(t)}^{\mathrm{r}}} a_{k(t)j}\varepsilon_j(t)\boldsymbol{q}_{k(t)}^{\mathrm{T}}(t)\boldsymbol{q}_{k(t)}(t).\end{aligned} \tag{15}$$

Considering (3) and (10), we can rewrite (15) as

$$\dot{W}_1(t) \leqslant \frac{1}{2}\sum_{j\in\mathscr{N}_{k(t)}^{\mathrm{r}}} a_{k(t)j}\varepsilon_{k(t)}(t)\left(2 - \varepsilon_{k(t)}^2(t) - \varepsilon_j^2(t)\right) - \sum_{j\in\mathscr{N}_{k(t)}^{\mathrm{r}}} a_{k(t)j}\varepsilon_j(t)\left(1 - \varepsilon_{k(t)}^2(t)\right),$$

which further leads to

$$\dot{W}_1(t) \leqslant -\sum_{j\in\mathscr{N}_{k(t)}^{\mathrm{r}}} a_{k(t)j}\left(\varepsilon_j(t) - \varepsilon_{k(t)}(t)\right) - \frac{1}{2}\sum_{j\in\mathscr{N}_{k(t)}^{\mathrm{r}}} a_{k(t)j}\varepsilon_{k(t)}(t)\left(\varepsilon_{k(t)}(t) - \varepsilon_j(t)\right)^2 \leqslant 0. \tag{16}$$

Moreover, considering (3), we can rewrite (11) as

$$W_1(t) = 2 - 2\varepsilon_{k(t)}(t) = 2 - 2\varepsilon_*(t) \tag{17}$$

for all $t \geqslant 0$.

In view of (16) and (17), it is obtained that $\dot{\varepsilon}_*(t) \geqslant 0$ holds if $\varepsilon_*(t) \geqslant 0$. This, together with the condition $\varepsilon_*(0) \geqslant 0$, leads to $\dot{\varepsilon}_*(t) \geqslant 0$ for almost all $t$ and $\varepsilon_*(t) \geqslant 0$ for all $t$ according to the so-called real induction or continuous induction in [27].

*2)* Based on the above result, we can further derive that $\varepsilon_*(t) \geqslant \varepsilon_*(T) > \varepsilon_*(0)$, $\forall t \geqslant T$ where $T \geqslant 0$ is some time instant. To prove this, we can apply the reduction to absurdity. Considering $\dot{\varepsilon}_*(t) \geqslant 0$ for almost all $t$ from 1), we suppose, on the contrary, $\varepsilon_*(t) = \varepsilon_{k(t)}(t) = \varepsilon_*(0) = 0$ holds for any $t \geqslant 0$, which implies that the scalar parts of all neighbors of the rigid body $k(t)$ are equal to 0 according to (16) and (17). Considering the strongly connected property, we can obtain that $\varepsilon_i(t) = 0$, $\forall i \in \mathscr{N}^{\mathrm{r}}$, $\forall t \geqslant 0$, which is a contradiction with the initial condition i1). Thus, it is obvious that $\varepsilon_*(t) \geqslant \varepsilon_*(T) > \varepsilon_*(0) = 0$, $\forall t \geqslant T$.

In addition, it is worth highly emphasizing that $\varepsilon_*(t) \leqslant 1$ always holds, which, together with that $\dot{\varepsilon}_*(t) \geqslant 0$ holds for almost all $t$, implies that $\lim_{t\to\infty}\varepsilon_*(t) = C_1$. Considering that $\varepsilon_*(t) \geqslant \varepsilon_*(T) > 0$, $\forall t \geqslant T$, we arrive at $C_1 > 0$.

To derive that $\lim_{t\to\infty} \varepsilon_i(t) = C_1$ holds for all $i \in \mathscr{N}^{\mathrm{r}}$, we invoke the reduction to absurdity again. With considering the definition of $\varepsilon_*(t)$, we suppose that $\limsup_{t\to\infty} \varepsilon_i(t) > C_1$, $\exists i \in \mathscr{N}^{\mathrm{r}}$ holds. It is obtained that $\varepsilon_*(t)$ won't remain at $C_1$ by virtue of (16) and (17) because each node has the neighbor nodes according to the strong connectivity, which makes a contradiction with $\lim_{t\to\infty} \varepsilon_*(t) = C_1$. As a consequence, it is obtained that $\lim_{t\to\infty} \varepsilon_i(t) = C_1$, $\forall i \in \mathscr{N}^{\mathrm{r}}$. The proof of Lemma 3 is complete. □

If the initial condition i1) is met, we might adopt Lemma 3 to analyze the evolution of $\varepsilon_i(t)$. However, it does not work when i2) is met because $\varepsilon_i(t) = 0$ holds for all $i \in \mathscr{N}^{\mathrm{r}}$ and $t \geqslant 0$ according to (2), (5), and (6). To address this problem, we can construct an appropriate coordinate transformation, which is stated in detail in the following lemma.

**Lemma 4.** *Let the initial condition i2) hold for $\mathscr{G}^{\mathrm{r}}$. There exists some $\boldsymbol{u} \triangleq [\varepsilon_u, \boldsymbol{q}_u^{\mathrm{T}}]^{\mathrm{T}} \in \mathbb{S}^3$ denoting the transformation quaternion of another inertial frame $\mathscr{F}_{\hat{I}}$ with respect to $\mathscr{F}_I$ such that attitudes of all rigid bodies can be expressed as*

$$
\begin{aligned}
\hat{\varepsilon}_i(t) &= \varepsilon_i(t)\varepsilon_u + \boldsymbol{q}_i^{\mathrm{T}}(t)\boldsymbol{q}_u \\
\hat{\boldsymbol{q}}_i(t) &= \varepsilon_u \boldsymbol{q}_i(t) - \varepsilon_i(t)\boldsymbol{q}_u + \boldsymbol{q}_i^{\times}(t)\boldsymbol{q}_u,
\end{aligned} \tag{18}
$$

*for all $i \in \mathscr{N}^{\mathrm{r}}$ in $\mathscr{F}_{\hat{I}}$ with the initial quaternions satisfying $\hat{\varepsilon}_i(0) \geqslant 0$, $\forall i \in \mathscr{N}^{\mathrm{r}}$ and $\max_{i\in\mathscr{N}^{\mathrm{r}}} \hat{\varepsilon}_i(0) > 0$.*

*Proof.* For $t = 0$, $\hat{\varepsilon}_i(0) = \boldsymbol{q}_i^{\mathrm{T}}(0)\boldsymbol{q}_u$ holds according to the initial condition i2). To prove the existence of $\boldsymbol{u}$, we can divide the analysis into two parts with considering $\boldsymbol{q}_i(0)$.

*Part 1: $q_i^{(3)}(0) > 0$, $\exists i \in \mathscr{N}^{\mathrm{r}}$.* It is obvious that the candidate $\boldsymbol{u}$ can be chosen as $[0, 0, 0, 1]^{\mathrm{T}}$ to provide $\hat{\varepsilon}_i(0) \geqslant 0$ for all $i \in \mathscr{N}^{\mathrm{r}}$, where not all $\hat{\varepsilon}_i(0)$ are zero.

*Part 2: $q_i^{(3)}(0) = 0$, $\forall i \in \mathscr{N}^{\mathrm{r}}$.* Under this condition, we can have the observation that $\boldsymbol{u}$ can be $[0, 1, 0, 0]^{\mathrm{T}}$ if $q_i^{(1)}(0) = 1$, $\forall i \in \mathscr{N}^{\mathrm{r}}$ and $[0, 0, 1, 0]^{\mathrm{T}}$, otherwise.

From the above discussions, we can obtain the existence of $\boldsymbol{u}$ satisfying (18). The proof of Lemma 4 is complete. □

**Remark 2.** *It is shown that Lemma 4 bridges the gap between two initial conditions i1) and i2) through a coordinate transformation. In addition, it is worth mentioning that the coordinate transformation does not change multiplicative quaternion errors and thus is reasonable in the analysis of the attitude synchronization.*

**Remark 3.** *Note that, in the analysis, we only need to ensure the existence of $\mathbf{u}$ in Lemma 4, and do not require its value in the design of the protocol. Moreover, although the analysis is based on the information of the networked rigid bodies, the synchronization protocol for each rigid body is designed only with the attitudes of its nearest neighbors and itself.*

Further, it is worth emphasizing that

$$\hat{\varepsilon}_{i,j}(t) = \varepsilon_{i,j}(t) \quad \text{and} \quad \hat{\boldsymbol{q}}_{i,j}(t) = \boldsymbol{q}_{i,j}(t) \tag{19}$$

hold for any rigid body $i$ and $j$ according to [23], where

$$\begin{aligned} \hat{\varepsilon}_{i,j}(t) &\triangleq \hat{\varepsilon}_i(t)\hat{\varepsilon}_j(t) + \hat{\boldsymbol{q}}_i^{\mathrm{T}}(t)\hat{\boldsymbol{q}}_j(t) \\ \hat{\boldsymbol{q}}_{i,j}(t) &\triangleq \hat{\varepsilon}_j(t)\hat{\boldsymbol{q}}_i(t) - \hat{\varepsilon}_i(t)\hat{\boldsymbol{q}}_j(t) + \hat{\boldsymbol{q}}_i^{\times}(t)\hat{\boldsymbol{q}}_j(t), \end{aligned} \tag{20}$$

denote the scalar part and the vector part of the multiplicative quaternion error expressed in $\mathscr{F}_{\hat{I}}$, respectively. As a result, attitude kinematics can be expressed in $\mathscr{F}_{\hat{I}}$ as

$$\begin{aligned} \dot{\hat{\varepsilon}}_i(t) &= -\frac{1}{2}\hat{\boldsymbol{q}}_i^{\mathrm{T}}(t)\boldsymbol{\omega}_i(t) \\ \dot{\hat{\boldsymbol{q}}}_i(t) &= \frac{1}{2}\left(\hat{\boldsymbol{q}}_i^{\times}(t) + \hat{\varepsilon}_i(t)I\right)\boldsymbol{\omega}_i(t) \end{aligned} \tag{21}$$

according to (2), (6), (18), and (19). It is worth noticing that $\boldsymbol{\omega}_i(t)$ in (21) is same with that in (2) because (19) holds. This shows us that the angular velocity input computed by (6) in $\mathscr{F}_I$ can be directly used in kinematics expressed in $\mathscr{F}_{\hat{I}}$.

### *B. Evolution of the vector parts of quaternions*

In addition to the above results of Lemmas 3 and 4, to prove the global attitude synchronization theorem of $\mathscr{G}^{\mathrm{r}}$ based on multiplicative quaternion errors, we still need to recall the following lemma.

**Lemma 5** [28]. *(Lemma 8.2, Barbalat's lemma) Suppose that $f:\mathbb{R}\to\mathbb{R}$ is uniformly continuous on $[0,\infty)$ and that $\lim_{t\to\infty}\int_0^t f(\tau)\mathrm{d}\tau$ exists. Then $\lim_{t\to\infty} f(t) = 0$.*

With Lemmas 3-5, we are in a position to present the global synchronization result of the strongly connected topologies $\mathscr{G}^{\mathrm{r}}$ based on multiplicative quaternion errors in the following theorem.

**Theorem 1.** *Let $\mathscr{G}^{\mathrm{r}}$ be strongly connected. For any initial physical attitudes of networked rigid bodies, if the protocol* (6) *is applied, then global attitude synchronization over $\mathscr{G}^{\mathrm{r}}$ can be achieved.*

*Proof.* According to (9), we only need to consider two kinds of initial conditions, i.e., i1) and i2).

**Step 1): Coordinate transformation.** If the initial condition i2) (i.e., $\varepsilon_i(0) = 0$ and $\boldsymbol{q}_i(0) \in S_+, \forall i \in \mathscr{N}^{\mathrm{r}}$) is satisfied, there must exist some coordinate transformation which can be constructed by employing Lemma 4 such that $\hat{\varepsilon}_i(0) \geqslant 0, \forall i \in \mathscr{N}^{\mathrm{r}}$ and $\max_{i\in\mathscr{N}^{\mathrm{r}}} \hat{\varepsilon}_i(0) > 0$. Similarly, if the other condition i1) is met, $\boldsymbol{u}$ can be chosen as $[1,0,0,0]^{\mathrm{T}}$ such that $\hat{\varepsilon}_i(0) = \varepsilon_i(0)$ and $\hat{\boldsymbol{q}}_i(0) = \boldsymbol{q}_i(0)$ hold. This implies that there exists some coordinate transformation in any situation. This fact, together with (21) and (19), shows that it is reasonable to analyze the attitude synchronization based on $\hat{\varepsilon}_i(t)$, $\hat{\boldsymbol{q}}_i(t)$, and $\hat{\boldsymbol{q}}_{i,j}(t)$.

**Step 2): Attitude synchronization analysis.** We can design a positive energy function as

$$W_2(t) = \sum_{i\in\mathscr{N}^{\mathrm{r}}} \left(\|\hat{\boldsymbol{q}}_i(t)\|^2 + (\hat{\varepsilon}_i(t) - 1)^2\right). \tag{22}$$

It is worth emphasizing that (22) can be reformulated into $W_2(t) = \sum_{i\in\mathscr{N}^{\mathrm{r}}} (2 - 2\hat{\varepsilon}_i(t))$ in terms of (3). Note also that from Lemma 3 and the existence of the coordinate transformation in Step 1), we obtain

$$\lim_{t\to\infty} \hat{\varepsilon}_i(t) = \hat{C}_1, \quad \forall i \in \mathscr{N}^{\mathrm{r}}, \tag{23}$$

where $\hat{C}_1 \in \mathbb{R}$ is some positive constant, which thus implies

$$\lim_{t\to\infty} \sum_{i\in\mathscr{N}^{\mathrm{r}}} \sum_{j\in\mathscr{N}_i^{\mathrm{r}}} a_{ij} \left(\hat{\varepsilon}_i(t) - \hat{\varepsilon}_j(t)\right) = 0 \tag{24}$$

and

$$\lim_{t\to\infty} W_2(t) = 2N^{\mathrm{r}} - 2N^{\mathrm{r}}\hat{C}_1. \tag{25}$$

It then follows from (25) that $\lim_{t\to\infty} \int_0^t \dot{W}_2(\tau)\mathrm{d}\tau$ exists because of the natural boundedness of $W_2(0)$. It is worth noticing that this fact is critically important for the subsequent analysis.

With the properties of $(\cdot)^{\times}$ in Section I and (21), it is obtained that

$$\dot{W}_2(t) = \sum_{i\in\mathscr{N}^{\mathrm{r}}} \hat{\boldsymbol{q}}_i^{\mathrm{T}}(t)\boldsymbol{\omega}_i(t). \tag{26}$$

With the similar analysis steps to (13)-(16) in the proof of Lemma 3, it can be shown that (26) leads to

$$\dot{W}_2(t) = \sum_{i\in\mathscr{N}^{\mathrm{r}}}\sum_{j\in\mathscr{N}_i^{\mathrm{r}}} a_{ij}\left(\hat{\varepsilon}_i(t)-\hat{\varepsilon}_j(t)\right) - \frac{1}{2}\sum_{i\in\mathscr{N}^{\mathrm{r}}}\sum_{j\in\mathscr{N}_i^{\mathrm{r}}} a_{ij}\hat{\varepsilon}_i(t)\left(\hat{\varepsilon}_i(t)-\hat{\varepsilon}_j(t)\right)^2 - \frac{1}{2}\sum_{i\in\mathscr{N}^{\mathrm{r}}}\sum_{j\in\mathscr{N}_i^{\mathrm{r}}} a_{ij}\hat{\varepsilon}_i(t)\left\|\hat{\boldsymbol{q}}_i(t)-\hat{\boldsymbol{q}}_j(t)\right\|^2. \tag{27}$$

By noticing the definition of $\hat{\varepsilon}_{i,j}(t)$ in (20) and inserting (24) into (27), we can deduce

$$\dot{W}_2(t) = \sum_{i\in\mathscr{N}^{\mathrm{r}}}\sum_{j\in\mathscr{N}_i^{\mathrm{r}}} a_{ij}\left(\hat{\varepsilon}_i(t)-\hat{\varepsilon}_j(t)\right) + \sum_{i\in\mathscr{N}^{\mathrm{r}}}\sum_{j\in\mathscr{N}_i^{\mathrm{r}}} a_{ij}\hat{\varepsilon}_i(t)\left(\hat{\varepsilon}_{i,j}(t)-1\right). \tag{28}$$

To use Lemma 5, let us check the uniform continuity of $\dot{W}_2(t)$ with the boundedness of $\ddot{W}_2(t)$. From (28), we obtain

$$\ddot{W}_2(t) = \sum_{i\in\mathscr{N}^{\mathrm{r}}}\sum_{j\in\mathscr{N}_i^{\mathrm{r}}} a_{ij}\left(\dot{\hat{\varepsilon}}_i(t)\left(\hat{\varepsilon}_{i,j}(t)-1\right)+\hat{\varepsilon}_i(t)\dot{\hat{\varepsilon}}_{i,j}(t)\right) + \sum_{i\in\mathscr{N}^{\mathrm{r}}}\sum_{j\in\mathscr{N}_i^{\mathrm{r}}} a_{ij}\left(\dot{\hat{\varepsilon}}_i(t)-\dot{\hat{\varepsilon}}_j(t)\right). \tag{29}$$

With invoking (19) and $\varepsilon_{i,j}^2(t)+\boldsymbol{q}_{i,j}^{\mathrm{T}}(t)\boldsymbol{q}_{i,j}(t)=1$ again, we have $\left\|\hat{\boldsymbol{q}}_{i,j}(t)\right\|\leqslant 1$ and $\left|\hat{\varepsilon}_{i,j}(t)\right|\leqslant 1$. This fact, together with (6), implies $\|\boldsymbol{\omega}_i(t)\|\leqslant N^{\mathrm{r}}-1$. Now according to (21), it is obvious that

$$\left|\dot{\hat{\varepsilon}}_i(t)\right| \leqslant \frac{1}{2}\left(N^{\mathrm{r}}-1\right) \text{ and } \left\|\dot{\hat{\boldsymbol{q}}}_i(t)\right\| \leqslant \left(N^{\mathrm{r}}-1\right). \tag{30}$$

As a consequence of (20) and (30), we have

$$\left|\dot{\hat{\varepsilon}}_{i,j}(t)\right| \leqslant 3\left(N^{\mathrm{r}}-1\right). \tag{31}$$

By noticing $\left|\hat{\varepsilon}_{i,j}(t)-1\right|\leqslant 2$ and $\left|\hat{\varepsilon}_i(t)\right|\leqslant 1$ and considering (29)-(31), we obtain

$$\left|\ddot{W}_2(t)\right| \leqslant 5\left(N^{\mathrm{r}}-1\right)^3, \tag{32}$$

which implies that $\ddot{W}_2(t)$ is bounded and thus ensures that $\dot{W}_2(t)$ is uniformly continuous.

According to (22), (25), and (32), it follows that $\dot{W}_2(t)$ is further ensured to approach zero from Lemma 5, namely,

$$\lim_{t\to\infty}\dot{W}_2(t)=0. \tag{33}$$

As discussed in Section III, the coupling problem from the products of the scalar parts and vector parts of quaternions makes the synchronization difficult to analyze. In other words, due to the existence of $\hat{\varepsilon}_i(t)$, the synchronization can not be directly derived only through (24), (28), and (33). Thanks to (23) from Lemma 3, it follows that $\lim_{t\to\infty}\hat{\varepsilon}_{i,j}(t)=1$ and $\lim_{t\to\infty}\hat{\boldsymbol{q}}_{i,j}(t)=\mathbf{0}$ if $a_{ij}\neq 0$. According to the definition of strongly connected topologies, we can draw that

$$\lim_{t\to\infty}\hat{\varepsilon}_{i,j}(t)=1 \text{ and } \lim_{t\to\infty}\hat{\boldsymbol{q}}_{i,j}(t)=\mathbf{0}, \quad \forall i\neq j\in\mathscr{N}^{\mathrm{r}},$$

which further implies that

$$\lim_{t\to\infty}|\hat{\varepsilon}_i(t)-\hat{\varepsilon}_j(t)|=0$$

$$\lim_{t\to\infty}\left\|\hat{\boldsymbol{q}}_i(t)-\hat{\boldsymbol{q}}_j(t)\right\|=0$$

for all $i,j\in\mathscr{N}^{\mathrm{r}}$ according to (6), (19), (21), and Lemma 3.

**Step 3): Inverse coordinate transformation.** Because there exists a coordinate transformation constructed in Step 1, the corresponding inverse transformation must exist. Obviously, attitude synchronization over the strongly connected topology $\mathscr{G}^{\mathrm{r}}$ can be obtained according to (2). In conclusion, with the protocol (6), the global synchronization over $\mathscr{G}^{\mathrm{r}}$ can be achieved. The proof of Theorem 1 is complete. □

**Remark 4.** *Another fact worth highlighting is that the global synchronization result based on the double-energy-function method consisting of Lemma 3 and Theorem 1 is entirely different from the existing works [17]–[20] aiming to achieve non-global synchronization. This analysis framework consequently relaxes the initial condition and extends the applications of multiplicative quaternion error-based protocols.*

## V. Synchronization of all rigid bodies

In this section, the attitude evolution of all rigid bodies, including root-related and non-root-related rigid bodies, associated with the quasi-strongly connected graph $\mathscr{G}$ are treated together with the double-energy-function method.

Before the analysis of attitude evolution, we are required to consider the following initial conditions of all rigid bodies associated with $\mathscr{G}$.

ii1) $\left[\varepsilon_i(0),\boldsymbol{q}_i^{\mathrm{T}}(0)\right]^{\mathrm{T}}\in S_1$, $\exists i\in\mathscr{N}^{\mathrm{r}}$ and $\left[\varepsilon_i(0),\boldsymbol{q}_i^{\mathrm{T}}(0)\right]^{\mathrm{T}}\in S_1\cup S_2$, $\forall i\in\mathscr{N}$;

ii2) $\left[\varepsilon_i(0),\boldsymbol{q}_i^{\mathrm{T}}(0)\right]^{\mathrm{T}}\in S_2$, $\forall i\in\mathscr{N}^{\mathrm{r}}$ and $\left[\varepsilon_i(0),\boldsymbol{q}_i^{\mathrm{T}}(0)\right]^{\mathrm{T}}\in S_1\cup S_2$, $\forall i\in\mathscr{N}$.

For ii1), we can develop the following result under the quasi-strongly connected topologies, which is similar to Lemma 3 in Section IV-A.

**Lemma 6.** *Let $\mathscr{G}$ be quasi-strongly connected. If the initial condition ii1) is met and the protocol* (6) *is applied, then the attitude synchronization over $\mathscr{G}$ can be achieved.*

*Proof.* From the initial condition ii1), we can conclude that $\varepsilon_i(0) \geqslant 0$, $\forall i \in \mathscr{N}$ and $\max_{i\in\mathscr{N}^{\mathrm{r}}} \varepsilon_i(0) > 0$. To proceed, we can divide the analysis into three steps as follows.

**Step 1): The behaviors of $\varepsilon_i(t)$ and $\boldsymbol{q}_i(t)$ for all $i \in \mathscr{N}^{\mathrm{r}}$.** From the definition of $\mathscr{N}^{\mathrm{r}}$, it is obtained that $\mathscr{G}$ has paths from $\forall i \in \mathscr{N}^{\mathrm{r}}$ to $\forall j \in \mathscr{N}^{\mathrm{nr}}$, whereas $\mathscr{G}$ has none paths from $\forall j \in \mathscr{N}^{\mathrm{nr}}$ to $\forall i \in \mathscr{N}^{\mathrm{r}}$, which shows that the attitude behavior of $\forall i \in \mathscr{N}^{\mathrm{r}}$ can be analyzed without considering $\forall j \in \mathscr{N}^{\mathrm{nr}}$. This fact, together with the strong connectivity of $\mathscr{G}^{\mathrm{r}}$, shows that if the initial condition ii1) is met, then $\lim_{t\to\infty} \varepsilon_i(t) = C_1 > 0$ holds for all $i \in \mathscr{N}^{\mathrm{r}}$ according to Lemma 3. In addition, $\boldsymbol{q}_i(t)$ for all $i \in \mathscr{N}^{\mathrm{r}}$ also converge to some identical constant vector from Theorem 1.

**Step 2): The behavior of $\varepsilon_\star(t) \triangleq \min_{i\in\mathscr{N}} \varepsilon_i(t)$.** With the analogous analysis method to that of Lemma 3, we can design the following energy function

$$V(\varepsilon_\star(t)) = 2 - 2\varepsilon_\star(t), \tag{34}$$

which, together with the same analysis as that of Lemma 3, show us that $\lim_{t\to\infty} \varepsilon_\star(t)$ exists. Furthermore, it is not hard for us to obtain that $\lim_{t\to\infty} \varepsilon_\star(t) = C_1$ holds with the reduction to absurdity and the result of Step 1), which implies

$$0 < C_1 \leqslant \liminf_{t\to\infty} \min_{i\in\mathscr{N}^{\mathrm{nr}}} \varepsilon_i(t). \tag{35}$$

**Step 3): The behaviors of $\varepsilon_i(t)$ and $\boldsymbol{q}_i(t)$ for all $i \in \mathscr{N}$.** In order to prove the convergences of $\varepsilon_i(t)$ and $\boldsymbol{q}_i(t)$ for all $i \in \mathscr{N}$, we suppose, on the contrary, that the attitude synchronization is not achieved when time approaches infinity. As a consequence, thanks to (35), there exists some appropriate coordinate transformation such that

$$0 < \liminf_{t\to\infty} \min_{i\in\mathscr{N}^{\mathrm{nr}}} \tilde{\varepsilon}_i(t) < \tilde{C}_1 \tag{36}$$

holds, where $\tilde{\varepsilon}_i(t)$ denotes the scalar part of the quaternion of rigid body $i \in \mathscr{N}$ after some coordinate transformation and $\tilde{C}_1 \triangleq \lim_{t\to\infty} \tilde{\varepsilon}_i(t)$, $\forall i \in \mathscr{N}^{\mathrm{r}}$. This must not occur according to the analogue analysis of Step 2). Therefore, (4) can be achieved. The proof is complete. □

Moreover, we have the following lemma when ii2) is met.

**Lemma 7.** *Let the initial condition ii2) hold for $\mathscr{G}$. There exists some transformation quaternion $\boldsymbol{v} \in \{\boldsymbol{u}\}$ where $\boldsymbol{u}$ is defined in* (18) *such that attitudes of all rigid bodies can be expressed as*

$$\hat{\varepsilon}_i(t) = \varepsilon_i(t)\varepsilon_v + \boldsymbol{q}_i^{\mathrm{T}}(t)\boldsymbol{q}_v$$
$$\hat{\boldsymbol{q}}_i(t) = \varepsilon_v\boldsymbol{q}_i(t) - \varepsilon_i(t)\boldsymbol{q}_v + \boldsymbol{q}_i^{\times}(t)\boldsymbol{q}_v, \tag{37}$$

*for all $i \in \mathscr{N}$ with the initial quaternions satisfying $\hat{\varepsilon}_i(0) \geqslant 0$, $\forall i \in \mathscr{N}$ and* $\max_{i\in\mathscr{N}^{\mathbf{r}}} \hat{\varepsilon}_i(0) > 0$.

Note that Lemma 7 is totally different from Lemma 4. In Lemma 4, we are only required to show the existence of $\boldsymbol{u}$ such that $\hat{\varepsilon}_i(0) \geqslant 0$, $\forall i \in \mathscr{N}^{\mathbf{r}}$ and $\max_{i\in\mathscr{N}^{\mathbf{r}}} \hat{\varepsilon}_i(0) > 0$. However, with choosing some candidate $\boldsymbol{v} \in \{\boldsymbol{u}\}$ in Lemma 7, we should provide that $\hat{\varepsilon}_i(0) \geqslant 0$, $\forall i \in \mathscr{N}$, while ensuring that $\max_{i\in\mathscr{N}^{\mathbf{r}}} \hat{\varepsilon}_i(0) > 0$ holds. Obviously, the non-root-related rigid bodies makes the analysis of the existence of $\boldsymbol{v}$ more difficult. For example, if there exists a non-root-related rigid body with the positive scalar part and the negative vector part, three candidate $\boldsymbol{u}$ presented in Lemma 7 cannot be chosen as $\boldsymbol{v}$ here, otherwise $\hat{\varepsilon}_i(0) < 0$ appears. The detailed analysis is presented as follows.

*Proof of Lemma 7.* Considering the definition of the initial condition ii2), we naturally divide the analysis into four parts:

1. $q_i^{(3)}(0) > 0$, $\exists i \in \mathscr{N}^{\mathbf{r}}$, $\left[\varepsilon_i(0), \boldsymbol{q}_i^{\mathrm{T}}(0)\right]^{\mathrm{T}} \in S_2$, $\forall i \in \mathscr{N}^{\mathbf{nr}}$;
2. $q_i^{(3)}(0) > 0$, $\exists i \in \mathscr{N}^{\mathbf{r}}$, $\left[\varepsilon_i(0), \boldsymbol{q}_i^{\mathrm{T}}(0)\right]^{\mathrm{T}} \in S_1$, $\exists i \in \mathscr{N}^{\mathbf{nr}}$;
3. $q_i^{(3)}(0) = 0$, $\forall i \in \mathscr{N}^{\mathbf{r}}$, $\left[\varepsilon_i(0), \boldsymbol{q}_i^{\mathrm{T}}(0)\right]^{\mathrm{T}} \in S_2$, $\forall i \in \mathscr{N}^{\mathbf{nr}}$;
4. $q_i^{(3)}(0) = 0$, $\forall i \in \mathscr{N}^{\mathbf{r}}$, $\left[\varepsilon_i(0), \boldsymbol{q}_i^{\mathrm{T}}(0)\right]^{\mathrm{T}} \in S_1$, $\exists i \in \mathscr{N}^{\mathbf{nr}}$.

Next, we investigate these parts one by one to reveal the existence of $\boldsymbol{v}$.

***Part 1:*** It is not hard to find that the candidate $\boldsymbol{v}$ can be $[0, 0, 0, 1]^{\mathrm{T}}$ such that $\max_{i\in\mathscr{N}^{\mathbf{r}}} \hat{\varepsilon}_i(0) > 0$ and $\hat{\varepsilon}_i(0) \geqslant 0$, $\forall i \in \mathscr{N}$ according to (37).

***Part 2:*** With a scalar $\epsilon_1$ defined as

$$\epsilon_1 \triangleq \min\{\varepsilon_i(0) : i \in \mathscr{N}^{\mathbf{nr}} \text{ and } \varepsilon_i(0) > 0\},$$

we can present the candidate $\boldsymbol{v}$ as

$$\boldsymbol{v} = \left[\left(1 - \epsilon_1^2\right)^{\frac{1}{2}}, 0, 0, \epsilon_1\right]^{\mathrm{T}}.$$

It is noted that for $\left[\varepsilon_i(0), \boldsymbol{q}_i^{\mathrm{T}}(0)\right]^{\mathrm{T}} \in S_1$, $i \in \mathscr{N}^{\mathbf{nr}}$, we know that $\varepsilon_i(0) \geqslant \epsilon_1 > 0$ and $q_i^{(3)}(0) \geqslant -\left(1 - \epsilon_1^2\right)^{1/2}$ holds and thus obtain that $\hat{\varepsilon}_i(0) = \left(1 - \epsilon_1^2\right)^{1/2} \varepsilon_i(0) + \epsilon_1 q_i^{(3)}(0) \geqslant 0$. Moreover, the transformation of other quaternions is easy to handle and thus is omitted.

***Part 3:*** In this part, $\varepsilon_i(0) = 0$ always holds, which means that $\|\boldsymbol{q}_i(0)\| = 1, \forall i \in \mathscr{N}$. With following two definitions:

$$\epsilon_2 \triangleq \min\{q_i^{(1)}(0) : i \in \mathscr{N} \text{ and } q_i^{(3)}(0) = 0\}$$
$$\epsilon_3 \triangleq \min\{q_i^{(3)}(0) : i \in \mathscr{N}^{\mathrm{nr}} \text{ and } q_i^{(3)}(0) > 0\},$$

we can choose the candidate $\boldsymbol{v}$ as

$$\boldsymbol{v} = \left[0, \left(\frac{1+\epsilon_2}{2}\right)^{\frac{1}{2}} \epsilon_3, \left(\frac{1-\epsilon_2}{2}\right)^{\frac{1}{2}} \epsilon_3, \left(1-\epsilon_3^2\right)^{\frac{1}{2}}\right]^{\mathrm{T}},$$

which definitely guarantees that $\hat{\varepsilon}_i(0) \geqslant 0, \forall i \in \mathscr{N}$ and $\max_{i \in \mathscr{N}^{\mathrm{r}}} \hat{\varepsilon}_i(0) > 0$. To find a valid $\boldsymbol{v}$, refer to Fig. 4 where the blue arrowed line denotes the vector $\boldsymbol{q}_v$. Next, it follows that the analysis of the validness of above chosen $\boldsymbol{v}$.

From the definitions of $\epsilon_2$ and $\epsilon_3$, it follows that $-1 < \epsilon_2 \leqslant 1$ and $\epsilon_3 > 0$ always hold. For any $\left[\varepsilon_i(0), \boldsymbol{q}_i^{\mathrm{T}}(0)\right]^{\mathrm{T}}$ with $q_i^{(3)}(0) > 0$, we have $q_i^{(3)}(0) \geqslant \epsilon_3$ and $\|\boldsymbol{q}_i(0)\| = 1$ and thus

$$\hat{\varepsilon}_i(0) = \left(\frac{1+\epsilon_2}{2}\right)^{\frac{1}{2}} \epsilon_3 q_i^{(1)}(0) + \left(\frac{1-\epsilon_2}{2}\right)^{\frac{1}{2}} \epsilon_3 q_i^{(2)}(0) + \left(1-\epsilon_3^2\right)^{\frac{1}{2}} q_i^{(3)}(0)$$

leading to

$$\begin{aligned}
\hat{\varepsilon}_i(0) &\geqslant \left(\frac{1+\epsilon_2}{2}\right)^{\frac{1}{2}} \epsilon_3 q_i^{(1)}(0) + \left(\frac{1-\epsilon_2}{2}\right)^{\frac{1}{2}} \epsilon_3 q_i^{(2)}(0) + \left(1-\epsilon_3^2\right)^{\frac{1}{2}} \epsilon_3 \\
&\geqslant -\left(1-\epsilon_3^2\right)^{\frac{1}{2}} \epsilon_3 + \left(1-\epsilon_3^2\right)^{\frac{1}{2}} \epsilon_3 \\
&= 0.
\end{aligned}$$

For any $\left[\varepsilon_i(0), \boldsymbol{q}_i^{\mathrm{T}}(0)\right]^{\mathrm{T}}$ with $q_i^{(3)}(0) = 0$, we have $1 \geqslant q_i^{(1)}(0) \geqslant \epsilon_2 > -1$ and invoking $\|\boldsymbol{q}_i(0)\| = 1$, and thus obtain

$$\begin{aligned}
\hat{\varepsilon}_i(0) &= \left(\frac{1+\epsilon_2}{2}\right)^{\frac{1}{2}} \epsilon_3 q_i^{(1)}(0) + \left(\frac{1-\epsilon_2}{2}\right)^{\frac{1}{2}} \epsilon_3 q_i^{(2)}(0) \\
&\geqslant \left(\frac{1+\epsilon_2}{2}\right)^{\frac{1}{2}} \epsilon_3 \\
&> 0.
\end{aligned}$$

As a consequence, in this part, we have $\hat{\varepsilon}_i(0) \geqslant 0, \forall i \in \mathscr{N}$ and $\max_{i \in \mathscr{N}^{\mathrm{r}}} \hat{\varepsilon}_i(0) > 0$.

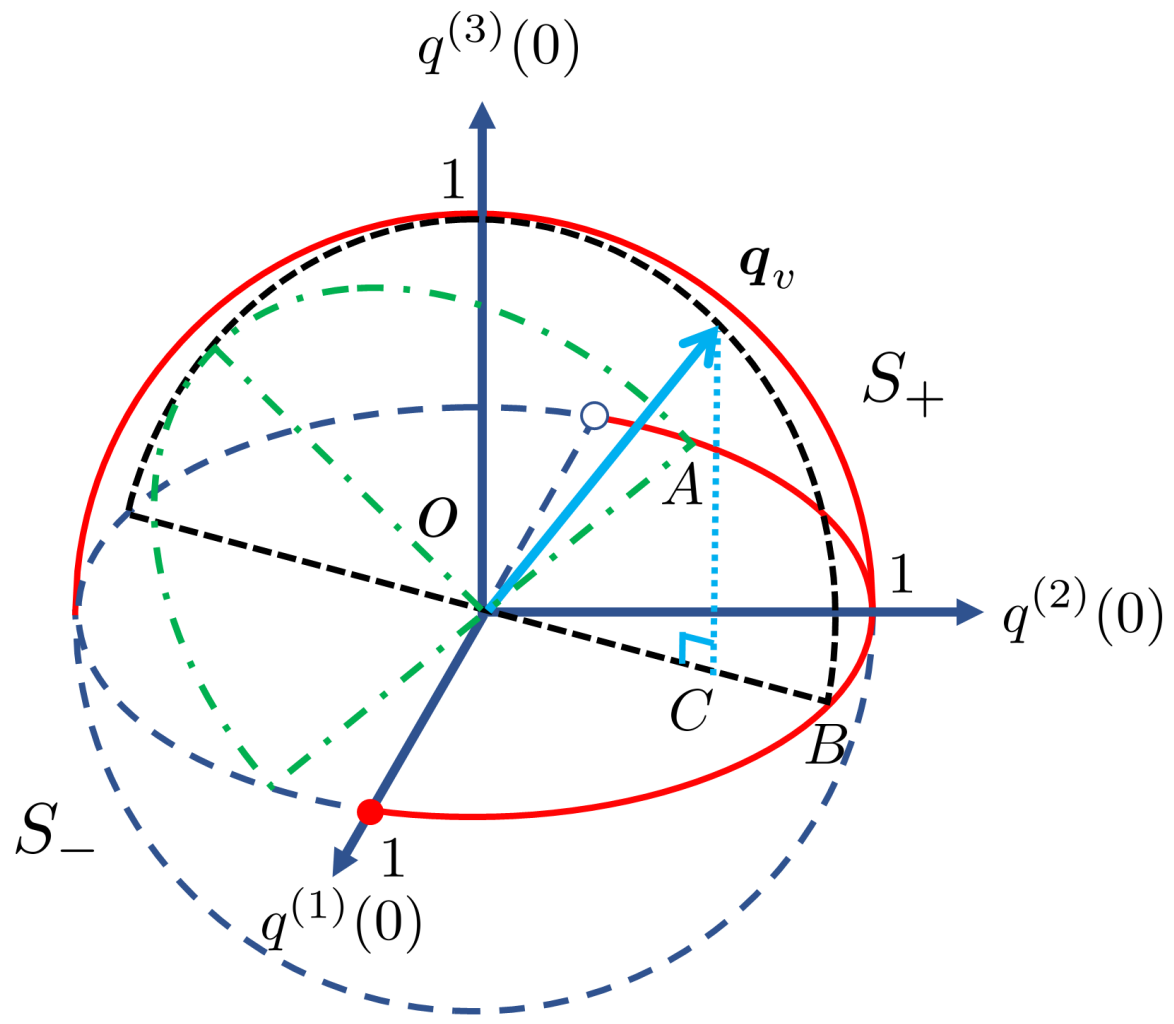

Fig. 4. The illustration of the existence of $\boldsymbol{q}_v \in S_+$. The coordinate of $A$ is $[\epsilon_2, \left(1-\epsilon_2^2\right)^{1/2}, 0]$, the coordinate of $B$ is $[\left(\left(1+\epsilon_2\right)/2\right)^{1/2}, \left(\left(1-\epsilon_2\right)/2\right)^{1/2}, 0]$, the coordinate of $C$ is $[\left(\left(1+\epsilon_2\right)/2\right)^{1/2}\epsilon_3, \left(\left(1-\epsilon_2\right)/2\right)^{1/2}\epsilon_3, 0]$, and the green dash-dotted line represents the plane perpendicular to $\boldsymbol{q}_v$.

***Part 4:*** We also have another three definitions:

$$\epsilon_4 \triangleq \min\{q_i^{(1)}(0) : i \in \mathscr{N},\ \varepsilon_i(0) = 0, \text{ and } q_i^{(3)}(0) = 0\}$$

$$\epsilon_5 \triangleq \min\{q_i^{(3)}(0) : i \in \mathscr{N}^{\mathbf{nr}},\ \varepsilon_i(0) = 0, \text{ and } q_i^{(3)}(0) > 0\}$$

$$\epsilon_6 \triangleq \min\{\varepsilon_i(0) : i \in \mathscr{N}^{\mathbf{nr}} \text{ and } \varepsilon_i(0) > 0\}.$$

With the similar analysis process to that in Part 3, we can prove that the candidate $\boldsymbol{v}$ consists of

$$\varepsilon_v = \left(1-\epsilon_6^2\right)^{\frac{1}{2}}$$

$$\boldsymbol{q}_v = \left[\left(\frac{1+\epsilon_4}{2}\right)^{\frac{1}{2}}\epsilon_5\epsilon_6,\ \left(\frac{1-\epsilon_4}{2}\right)^{\frac{1}{2}}\epsilon_5\epsilon_6,\ \left(1-\epsilon_5^2\right)^{\frac{1}{2}}\epsilon_6\right]^{\mathrm{T}}.$$

From the above analysis of Parts 1-4, we can arrive at the existence of $\boldsymbol{v}$ under the initial condition ii2). The proof of Lemma 7 is complete. □

As a consequence, the attitude synchronization problem with the quasi-strongly connected topologies based on multiplicative quaternion errors can be addressed with the help of Lemmas 6 and 7, which is organized in the following theorem.

**Theorem 2.** *For a group of $N$ rigid bodies whose attitudes are initially arbitrary, let the attitude synchronization protocol* (6) *be applied. Then the global attitude synchronization objective* (4) *can be achieved if and only if the communication topologies are quasi-strongly connected.*

*Proof. Sufficiency:* Obviously, according to (9), there exists some appropriate coordinate transformation quaternion $\boldsymbol{v}$ according to Lemma 7 such that $\hat{\varepsilon}_i(0) \geqslant 0, \forall i \in \mathscr{N}$ and $\max_{i \in \mathscr{N}} \hat{\varepsilon}_i(0) > 0$ hold. Consequently, it is concluded that the attitude synchronization in $\mathscr{F}_{\hat{I}}$ can be achieved according to Lemma 6 regardless of the initial attitudes if the topology graph is quasi-strongly connected. This also implies that the global attitude synchronization objective (4) can be achieved in $\mathscr{F}_I$. *Necessity:* For a time-invariant communication topology, if the quasi-strong connectivity is not satisfied, the synchronization can not be achieved [25]. The proof is complete. □

**Remark 5.** *Thanks to the proposed double-energy-function analysis method, global synchronization can be verified under quasi-strongly connected topologies. This method is different from these in [17]–[20], which are aimed at the attitude synchronization under strongly connected or undirected topologies. Clearly, this condition of topologies is more general and practical.*

**Remark 6.** *To clearly demonstrate the idea of the global synchronization performance, we only investigate the kinematics without regard to torques, which is also implemented in literature, e.g., [10], [29]. Furthermore, with the tracking of the desired angular velocity* (6)*, it is easy to simultaneously consider the kinematics and dynamics via the back stepping method or the dynamic surface control [14], [30].*

## VI. Simulations

To illustrate the validity of the proposed global attitude synchronization algorithm, two cases for one group consisting of five rigid spacecraft under different topological graphs are considered in this section.

*Case 1:* A strongly connected topology. The nonzero edge weights of the adjacency matrix $\mathscr{A}$ are given by $a_{15} = 1.0$, $a_{21} = 0.5$, $a_{32} = 0.8$, $a_{43} = 0.6$, and $a_{54} = 0.3$. The corresponding communication graph of the networked rigid spacecraft is illustrated in Fig. 1.

*Case 2:* A quasi-strongly connected topology. The nonzero edge weights of the adjacency matrix $\mathscr{A}$ are given by $a_{15} = 1.0$, $a_{24} = 0.5$, $a_{32} = 0.8$, $a_{43} = 0.6$, and $a_{54} = 0.3$. The corresponding communication graph of the networked rigid spacecraft is illustrated in Fig. 2.

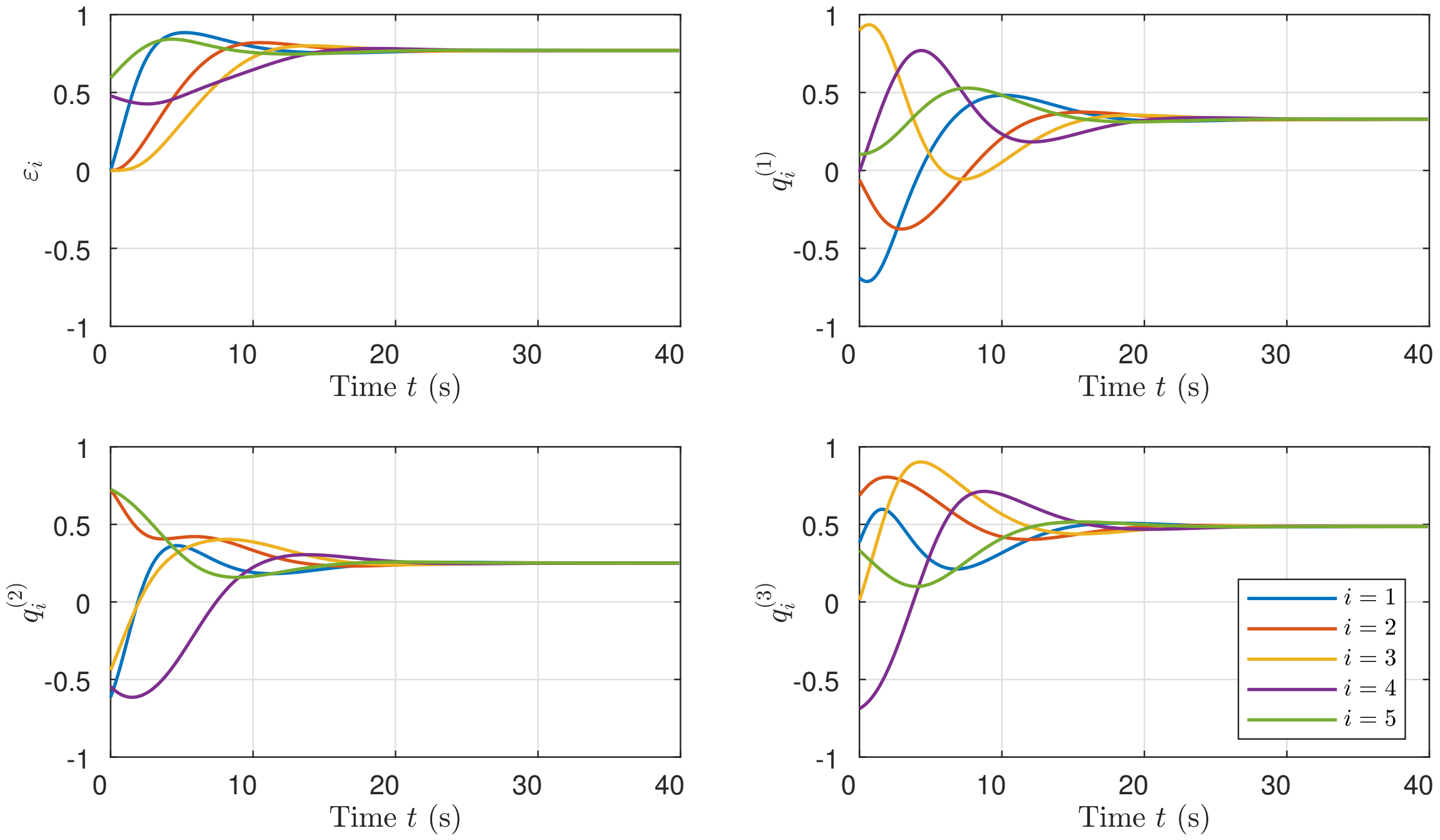

Fig. 5. Simulation results under strongly connected topologies (Case 1).

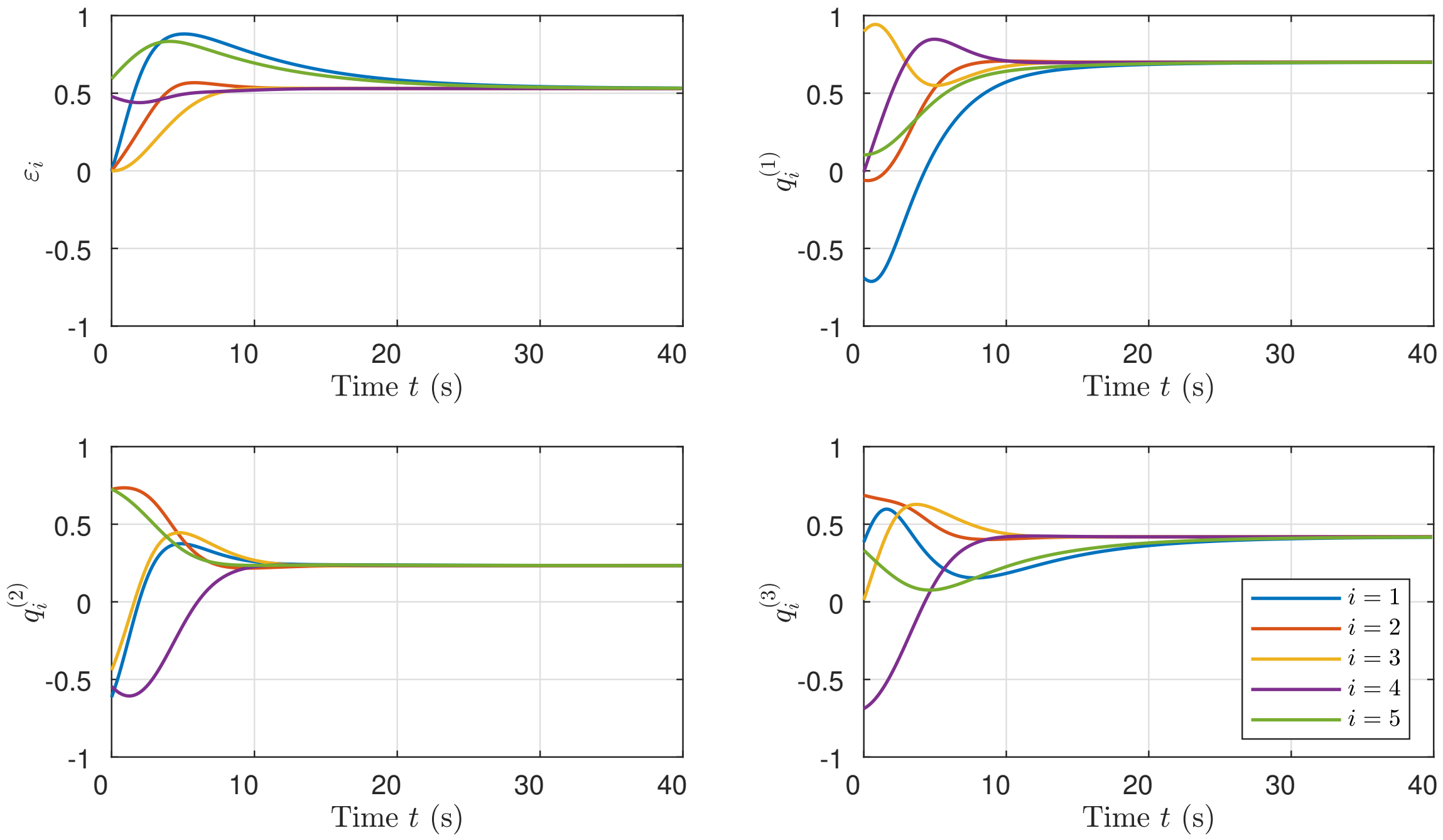

Fig. 6. Simulation results under quasi-strongly connected topologies (Case 2).

Moreover, the initial quaternions can be given arbitrarily. For the sake of simplicity, the initial quaternions of five rigid spacecraft are respectively set as

$$\begin{aligned}
\left[\varepsilon_1(0), \boldsymbol{q}_1(0)^{\mathrm{T}}\right]^{\mathrm{T}} &= [0, -0.6894, -0.6140, 0.3843]^{\mathrm{T}} \\
\left[\varepsilon_2(0), \boldsymbol{q}_2(0)^{\mathrm{T}}\right]^{\mathrm{T}} &= [0, -0.0602, 0.7248, 0.6863]^{\mathrm{T}} \\
\left[\varepsilon_3(0), \boldsymbol{q}_3(0)^{\mathrm{T}}\right]^{\mathrm{T}} &= [0, 0.8975, -0.4409, 0.0119]^{\mathrm{T}} \\
\left[\varepsilon_4(0), \boldsymbol{q}_4(0)^{\mathrm{T}}\right]^{\mathrm{T}} &= [0.4796, -0.0077, -0.5447, -0.6879]^{\mathrm{T}} \\
\left[\varepsilon_5(0), \boldsymbol{q}_5(0)^{\mathrm{T}}\right]^{\mathrm{T}} &= [0.5929, 0.1024, 0.7263, 0.3325]^{\mathrm{T}}.
\end{aligned}$$

Note also that the above initial condition with $\varepsilon_i(0) = 0$, $\exists i \in \mathscr{N}$ can not be handled with the analysis methods in [18] and [19] where the initial rotation matrices should be supposed to be positive definite, which is equivalent to that none of the scalar parts of quaternions can be zero.

The simulation results of Cases 1 and 2 are illustrated in Figs. 5 and 6, respectively. From the above cases, it is obvious that all quaternions converges to some constant values in both cases regardless of the initial attitude condition with the help of the synchronization protocol (6). Additionally, the simulation results in terms of $\varepsilon_*(t)$ can also be shown in the left-top subfigures of Figs. 5 and 6, where $\varepsilon_*(t)$ is non-decreasing along the time axis. These results effectively illustrate that the attitude synchronization objective can be achieved under strongly connected topologies or quasi-strongly connected topologies with any initial conditions, and verify the correctness and the rationality of the analysis framework.

In comparison to the existing works, e.g., [18], [19], it is shown that the initial condition can be extended from local scope to global scope under the analysis framework. Moreover, it has been demonstrated that the undirected graph could be relaxed to the strongly connected graph and furthermore the quasi-strongly connected graph, which can not be achieved with the existing analysis methods in the literature. Thus, it is revealed that the proposed double-energy-function method effectively expands the scope of application of multiplicative quaternion error-based synchronization protocol.

## VII. Conclusions

In the present work, the global attitude synchronization problem for networked rigid bodies has been investigated in the presence of directed topologies. To achieve this objective, we have involved multiplicative quaternion errors to design an attitude synchronization protocol for networked rigid bodies, rather than additive quaternion errors. Moreover, a new double-energy-function analysis method has been developed for global attitude synchronization. Specially, a novel energy function of the rigid body with the minimum scalar part, equipped with an ordering permutation technique and a coordinate transformation mechanism, has been proposed, which has been proved to guarantee the positive property of all the scalar parts when time tends to infinity in some newly defined inertial frame and thus relax the restricted initial condition. Based on this fact, another energy function of the networked rigid bodies has been proposed to prove that all rigid bodies are driven to approach the same attitude eventually. With the above double-energy-function method, global attitude synchronization has been rigorously analyzed based on Barbalat's lemma. Notably, the analysis method can be applied to quasi-strongly connected topologies, which has proved to be the sufficient and necessary condition for global attitude synchronization. Finally, for a practical example on a group of rigid spacecraft, various cases under strongly connected graph and quasi-strongly connected graph are illustrated, which have verified the global attitude synchronization performance. Based on this research, global attitude synchronization over the switching topologies will be one of our future works.

## Acknowledgment

The authors would like to acknowledge several important exchanges with Dr. Y. Wu from Beihang University.